# The working principle of magnetic resonance therapy

L. Brizhik[1], B. Zavan[2], E. Fermi[3],

[1]Bogolyubov Institute for Theoretical Physics
Metrolohichna Str., 14b, Kyiv 03680, Ukraine
E-mail: brizhik@bitp.kiev.ua

[2]Department of Histology, Microbiology, and Medical Biotechnology,
Università Degli Studi Di Padova, Via 8 Febbraio, 2 - 35122 Padova, Italy
E-mail: barbara.zavan@unipd.it

[3]THERESON S.p.A., Via Torri Bianche, 9 – Vimercate (MB), Italy,
E-mail: enrico.fermi@thereson.com

## ABSTRACT

In this paper we describe briefly the basic aspects of magnetic resonance therapy, registered as TMR® therapy. Clinical studies have shown that application of this therapy significantly accelerates wound healing and, in particular, healing of the diabetic foot disease. To understand the working principle of this therapy, we analyze relevant to it biological effects produced by magnetic fields. Based on these data, we show that there is a hierarchy of the possible physical mechanisms, which can produce such effects. The mutual interplay between the mechanisms can lead to a synergetic outcome delayed in time, which can affect the physiological state of the organism. In particular, we show that soliton mediated charge transport during the redox processes in living organisms is sensitive to magnetic fields, so that such fields can facilitate redox processes in particular, and can stimulate the healing effect of the organism in general. This and other non-thermal resonant mechanisms of the biological effects of magnetic fields are summarized as the working principle of the magnetic resonance therapy. We support our approach by some biological and histological data (both *in vitro* and *in vivo*) and, finally, by some clinical data.

## 1. Introduction

The increasing need in non-invasive and non-chemical (non-drug) therapies as alternative and/or supplementary to the drug-based medicine, stimulates the research of the biological effects, caused by electric and magnetic fields, static or alternating, and development of therapies based on application of these fields. In spite of the fact that organisms as the whole and the majority of biological tissues, in particular, are non-magnetic media, the fact of the biological effects of electromagnetic fields (EMF) is well established experimentally. The recent review of some of such effects, caused by electromagnetic fields of weak intensity in the broad interval of frequencies and intensities, can be found in (Brizhik, 2014a). Especially promising from the point of view of medical applications are low and extremely low frequency (ELF) fields which are widely used in various therapies to treat broad class of diseases. In particular, one of such medical applications is based on the therapy, administered with the devices produced by THERESON company (Italy). The corresponding method, called Therapeutic Magnetic Resonance TMR™, turned out to show positive results in wound healing, and, in particular, in treatment of diabetic foot disease and vascular ulcers (Book of Abstracts, 2013). The method TMR™ consists of exposing patients to low intensity Pulsating Electro-Magnetic Fields (PEMF), at specific patented protected shapes and low frequencies of pulses. Successful application of this therapy and further improvement of its efficiency require understanding of the responsible physical mechanisms and knowledge of its working principle, which is the aim of the present paper.

Below we give the brief review of the experimental data showing the biological effects of magnetic fields, describe the Therapeutic Magnetic Resonance TMR™ method and the main aspects of the processes of wound healing, provide the biological and clinical data of the results of this therapy, and, finally, develop the physical mechanisms, which can explain the action of this therapy. In an organism, which is highly nonlinear open system, these mechanisms are complementary, act simultaneously (at different time scales) and synergetically, although at different stages of the disease one or another can prevail. In Conclusion, summarizing these mechanisms, we formulate the working principle of the magnetic therapy.

## 2. The experimental background of magnetic therapies

Electric and magnetic fields often occur together, nevertheless, namely magnetic field (MF) can be responsible for the biological effects, because MF is not screened by the skin, is not absorbed by cell cytoplasm, and can easily penetrate deep into biological tissues. One of the recent reviews of the electromagnetic properties of biological tissues and electromagnetic phenomena at tissue interfaces can be found in (Lin, 2011). The review of the experimental data on the biological effects of the magnetic field has been reported in (Brizhik, 2014a). Here we update this review with the special emphasis on the therapeutic aspects of the impact of electromagnetic fields of low and extremely low frequencies on the living matter.

Low and extremely low frequency low-intensity electromagnetic fields are very promising from the point of view of their applications in medicine. The allostatic load on the organism exposed to such fields, is much less than in the case of high intensity and high frequency fields, even when we deal with the resonant mechanisms of the biological effects (Brizhik, 2014b). Exposed to LF or ELF low intensity MF, a cell (organ, system, organism) can adapt without drastic consequences and a transient perturbation can be followed by an adjustment of the system via the normal homeostatic machinery of the cells. In such a way the system is not moved far from its quasi-stationary state and can adapt to new conditions in a more natural way. Such adaptation is accompanied by the tendency to restore the healthy state.

Finally, it needs to be emphasized that no major, clinically important, side effects with the use of low frequency low-intensity MFs in numerous studies have been reported so far. The healthy system can even not respond to weak external stimuli at certain conditions. In terms of the theory of complex systems the 'healthy' and 'ill' states of an organism can be classified as two different attractors of the system in the multi-parameter phase space. These attractors correspond to quasi-stationary states, which have different free energies separated by the potential barrier. The transition between the two states requires the energy to overcome this potential barrier (Brizhik, 2014b). This is one of the reasons why the process of healing often occurs through the exacerbation phase (intensification of the disease) at the early stage of the treatment.

It is worth to recall that the biological effects, caused by MFs, are manifested on all levels of the hierarchy of the living matter. In particular, on the levels of macromolecules, cells, organs and systems as well as the whole organism (Brizhik, 2014a). The biological impact of the MF results from all biological pathways and processes, such as energy storage and transfer, charge (matter) transport, information transfer and exchange. Below we briefly describe some reliably established facts of the biological effects caused by MFs which, in our opinion, are important for their medical implementations.

It is clear that a system can be sensitive to MF if it possesses charge and/or magnetic momentum (spin). Therefore, the primary biological 'sensors' of external magnetic fields can be electrons, protons and ions, which are involved in all metabolic processes in living systems. These particles expire in a magnetic field **B** the Lorentz force, proportional to their charge, $q$, and velocity, $V$:

$$\boldsymbol{F} = q \boldsymbol{V} \boldsymbol{B}. \tag{1}$$

Ions with non-zero magnetic momentum and biological molecules containing such ions, can be also sensors of the magnetic field. It is known that ferrum ions have magnetic momentum and thus, react to the exposure to magnetic fields. Ferrum ions (iron) are included in hemoglobin in blood cells, and therefore, one should expect the effects of MFs on blood. Indeed, it has been proven experimentally that magnetic field exposure affects blood flow and state of blood vessels in the microvasculature (McKay, 2007), affects the microcirculation (Ohkubo, 2015), immune and anti-inflammatory responses (Balcavage, 2015). As it is evident from the spectral analysis of EEG, measurements of auditory-evoked potentials and reaction time (Lyskov, 1993a, 1993b), MF affects the functional state of the human brain and nervous system causing physiological effects. When the nervous system or the brain is disturbed, for instance, by EMFs, morphological, electrophysiological, and chemical changes can occur. Significant perturbations of these functions will inevitably lead to physiological and behavioral changes. Neurological effects of EMFs reported in literature, include changes in blood–brain barrier, morphology, electrophysiology, neurotransmitter functions, cellular metabolism, calcium efflux, responses to drugs that affect the nervous system, and behavior. Exposure to ELF magnetic and ELF-modulated radiofrequency can have also physiological and cognitive effects (Cook, 2006).

Oscillating magnetic fields can cause changes in conformations of macromolecules and their binding probabilities and, accordingly, generate biological effects. It is well known that the function of the macromolecule depends on (or even is determined by) its conformation. In particular, the efficiency of the protein as an enzyme depends on its conformation. For instance, it has been shown in (Bohr, 2000) that EMR affects the kinetics of conformational changes of the protein β-lactoglobulin and accelerates conformational changes in the direction toward the equilibrium state. Magnetic field can cause excitation and absorption of vibrational states of biological components such as microtubules, alpha-helical proteins, etc. (Habash, 2007; Brizhik,1999; Brizhik, 2002; Brizhik, 2009).

Basic biological mechanisms reviewed by many authors (see (Brizhik, 2014a; Lin, 2011; Habash, 2007; Markov, 2015; Leitgeb, 2011) and references therein) put some light on the effects of EMFs on biological systems, but we are still far from a deep understanding of this phenomenon. The site(s) and mechanisms of interaction of EMF and living matter still remain to be elaborated. Numerous hypotheses have been suggested, although some sceptics think that none is convincingly supported by experimental data. In our opinion, in a complex system, like an organism, one can hardly expect a single mechanism been responsible for any biological effect of any external stimulus. There is a hierarchy of matter organization in a living matter, and, respectively, a hierarchy of biological processes converting the primary effects of the external stimulus in a physiological response.

Nevertheless, magnetic fields found extensive applications in medicine in view of the beneficial effects discovered mainly phenomenologically. The importance of MF is widely known due to their use in the tools for medical diagnostics, such as magnetic resonance tomography. Various medical devices for magnetic therapy and diagnosis have been designed so far (Leitgeb, 2011). Different medical applications of EMFs are used extensively not only for body relaxation and wellness purposes, but also for treatment of various diseases. The history of magnetic therapy covers several decades at least (see, e.g., the reviews in (Habash, 2007; Markov, 2015; Andrew, 1993)). Here we use the term 'magnetic therapy' in a broad sense as therapy, based on the use of electromagnetic fields. These include the use of EMFs for pain relief and pain control, inhibition of cancer growth (Habash, 2007), for treatment of vascular fatigue, in plastic and reconstructive surgery (McKay, 2007), in cartilage and bone repair, in treatment of post-operative pain and edema (Markov, 1995) and in wound healing (Mayrovitz, 2015). Namely the application of MFs for wound healing will be studied in more details below.

Usually two types of therapies are used. In one of them the magnetic field is applied to upper (lower) limbs or the whole body is exposed to the magnetic field. This can cause a local or a systematic effect, when the benefit is obtained at sites of the direct exposition as in the case of wound healing, or at sites distant from the site of field application (Ericsson, 2004). In TMR™ the combination of the both approaches is used, which, in our opinion, gives the best result in particular because of the synergetics of the local and systematic effects (see below).

### 3. Using magnetic fields in wound healing

Treatment of chronic wounds by means of electric and electromagnetic fields (Vodovnik, 1992) has demonstrated positive results. Moreover, such therapies turned out to be useful for treatment of patients with painful diabetic polyneuropathy (low sensation of nerves caused by uncontrolled blood glucose levels) (Weintraub, 2003; Cieslar, 1995; Wróbel, 2008). Such therapies alone or in combination with existing clinical protocols can accelerate diabetic wound healing.

There are several factors that influence wound healing in a diabetic patient. These include high blood glucose levels, poor blood circulation, high risk of infections and some other. In particular, an elevated blood sugar level leads to stiffness of the arteries and narrowing of the blood vessels. The effects of this are far-reaching and include the origin of wounds as well as an increase of risk factors for proper wound healing. Narrowed blood vessels lead to decreased blood flow and oxygen to a wound. An elevated blood sugar level decreases the function of red blood cells that carry nutrients to the tissue. This lowers the efficiency of the white blood cells that fight infection. Without sufficient nutrients and oxygen, a wound heals slowly. Diabetes lowers the efficiency of the immune system, the body's defense system against infection. A high glucose level causes the immune cells to function ineffectively, which raises the risk of infection for the patient. Studies indicate that particular enzymes and hormones that the body produces in response to an elevated blood sugar, are responsible for negatively impacting the immune system. With a poorly functioning immune system, diabetics are at a higher risk for developing an infection. Infection raises many health concerns and also slows the overall healing process. Left untreated, infection can heighten the risk of developing gangrene, sepsis or a bone infection like osteomyelitis. According to statistics, diabetes is the primary reason for limb amputation in many countries.

### 3.1. Stages of wound healing

To elucidate the physical mechanisms of the magnetic therapies for wound healing, we need to summarize the biological processes, involved in the wound healing. The process of wound healing can be conditionally separated into four stages (phases) which are:

1. Hemostasis phase - formation of platelet plug, formation of a stable fibrin clot. It is a process which causes bleeding to stop, and involves blood changing from a liquid to a gel.

2. Inflammatory phase (substrate-preparation phase) which lasts 1--4 days. It involves inflammation process and migration of cells, such as platelets, neutrophils, lymphocytes, macrofages, endothelial progenitor cells (EPCs). Endothelial cells are thin flattened cells which line the inside surfaces of body cavities, blood vessels, and lymph vessels. EPCs circulate in the blood and can differentiate into endothelial cells.

3. Proliferation phase (collagen-building phase) which lasts 2--22 days. It involves cell proliferation, the extracellular matrix (ECM) synthesis, angiogenesis, re-epithelialization. It includes such cells as keratinocytes, endothelial cells, fibroblasts, macrofages, EPCs. Fibroblasts are cells that synthesize the extracellular matrix and collagen which play a critical role in wound healing.

4. Remodeling phase (maturation) which lasts 6--12 months and involves wound closure and contraction. Cell types involved are myofibroblasts, macrophages etc.

The impaired vascular supply associated with diabetes, leads to poor blood flow at the wound site impeding the optimal endogenous reparative response (Jeffcoat, 2003). In addition, neovascularization (formation of new blood vessels) is critical for granulation tissue formation and tissue regeneration in wound healing (Gurtner, 2008). The impaired angiogenic response that occurs in diabetes mellitus leads to hypoxia at the wound site (lack of adequate oxygen supply). Temporary hypoxia is requisite for normal wound healing. In the non-diabetic situation, hypoxia leads to activation of the transcription factor complex HIF-1alpha (Hypoxia inducible factor-1alpha), which leads to transcription of multiple genes required for successful wound healing. With diabetes, hyperglycaemia affects the stability and activation of HIF-1alpha. This suppresses platelet-derived growth factor, vascular endothelial growth factor and transforming growth factor-beta, which are required for angiogenesis, *in vitro* and *in vivo* wound healing.

### 3.2. Cellular biology study of the effects of TMR®

An extensive cellular biology study has been performed by Ferroni et al (Ferroni 2015) to ascertain the effects of TMR® stimulation on cell behavior. In particular, it has been demonstrated that PEMFs influences mitochondrial function, as it follows from reactive oxygen species (ROS) measurements. Reactive oxygen species are chemically reactive molecules containing oxygen, such as oxygen ions and peroxides, formed as a natural byproduct of the normal metabolism of oxygen and have important role in cell signaling and homeostasis. Under conditions of oxidative stress the cell accumulates ROS. When the culture cells, both fibroblasts and endothelial cells, are treated with TMR™, a well-defined decrease of ROS productions is revealed, as it is shown in Fig.1 a) and b). The fluorescence intensity is a measure of oxidative stress. The OxiSelect™ ROS Assay Kit has been used to measure fluorescence level of samples.

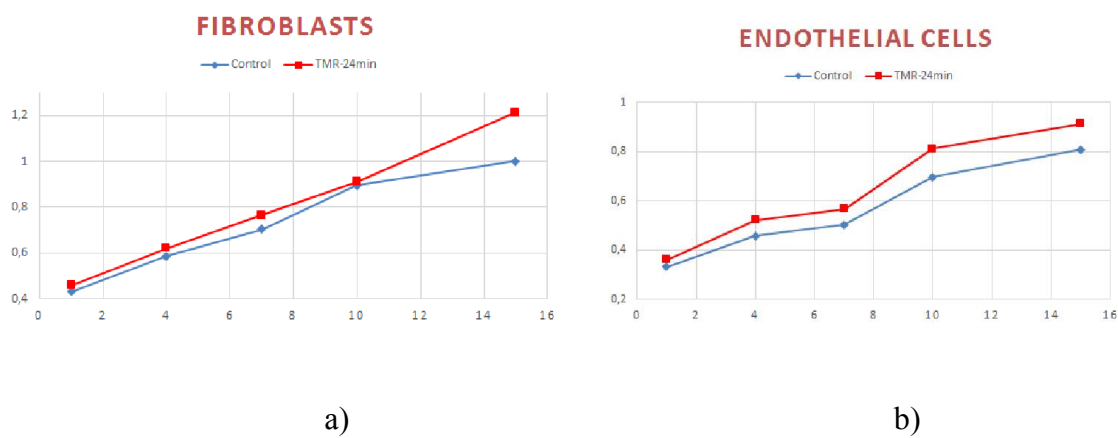

a) b)

Fig.1. Fluorescence intensity (a.u.) versus time (days) of a) fibroblasts and b) endothelial cells in control group (upper curves) and test group treated with PEMF (lower curves).

Moreover, TMR® exposure influences proliferation of normal endothelial cells, diabetic derived endothelial cells and fibroblast and endothelial as reported in Fig. 2. Our results confirm that a significant increase of time related to MTT value, is well evident for all cellular types. We remind that MTT test is a colorimetric assay for assessing cell metabolic activity and is based on measurement of MTT, a yellow tetrazole, which reduces to purple formazan in living cells. In solubilization solutions the insoluble purple formazan product dissolves into a colored solution.

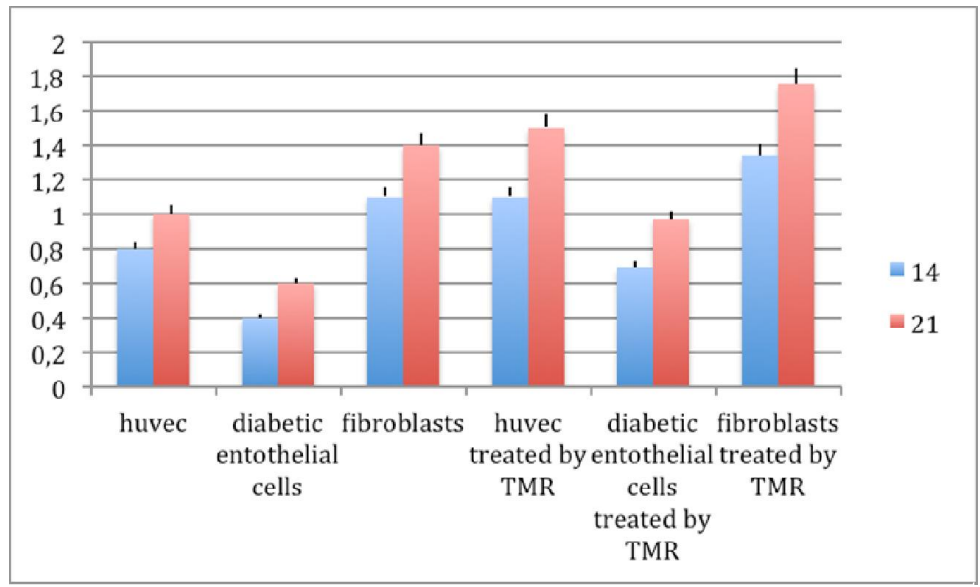

Fig. 2. Proliferation rate in control group and in a test group 14 and 21 days after exposure to PEMF of endothelial cells, diabetic derived endothelial cells and fibroblasts.

A subsequent more detailed analysis of extracellular matrix proteins as reported in fig. 3, showes that in the end of treatment with TMR the significant increase of the production of the most important proteins such as collagen type V, III; XIV and growth factors such as FGF 10/7/2 which are mainly involved in tissue repair process.

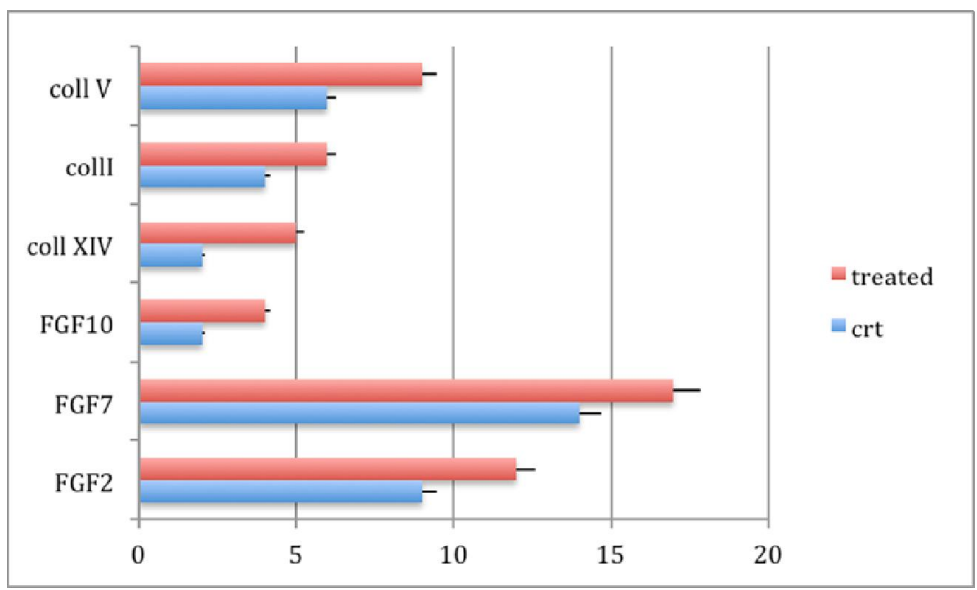

Fig. 3. Amount of different extracellular matrix proteins (in a.u.), involved in tissue repair process, in control tissues (lower histograms) and tissues, treated with TMR (in the end of treatment)

These markers strongly improve the quality of extracellular matrix supporting the growing of keratinocytes. In an *in vitro* model of a 3D artificial dermis, we demonstrated that if fibroblasts were previously treated with TMR, a better and faster growth of keratinocytes was observed. As reported in Fig. 4, a well defined multilayers of Keratinocytes are evident, indeed.

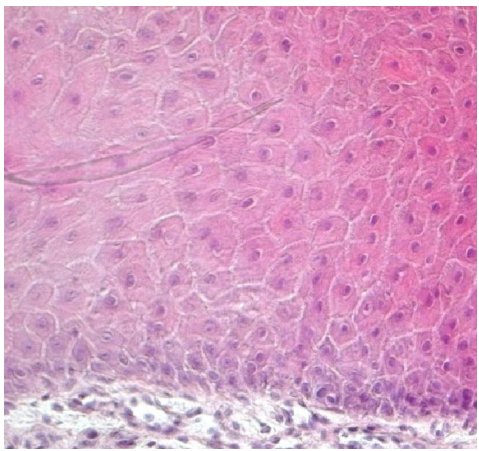

Fig.4. Histopathology analysis - microscopic examination - of a 3D artificial dermis, demonstrating keratinized dermis after treatment with TMR.

### 3.3. Clinical data on application of TMR® therapy

Also some clinical TMR® therapy data have been collected recently (Abbruzzese, 2015). The clinical experience in the application of TMR® therapy to diabetic foot ulcer treatment began in 2012. It includes several studies:

1. A study on surgical patients, conducted by Prof. Piaggesi (Cisanello Hospital – Pisa, Italy) in January - March 2012 with the 12 months follow-up.
2. A multi-center exploratory clinical study, conducted by three Italian clinical centers, in September 2012 – April 2014.
3. A bioptic «in vivo» study (a part of the above study) conducted in December 13 - March 2014, analyzed by University of Padua (Italy).
4. Remodeling phase (maturation) which lasts 6--12 months and involves wound closure and contraction. Cell types involved are myofibroblasts, macrophages etc.

An independent study on surgical patients conducted by Prof. A. Piaggesi (Abbruzzese, 2015) has shown statistically significant evidence of wound healing acceleration. In particular, he 90% healing rate at 6 months in the group treated with TMR® and 30% in group with standard treatment only (statistically significant; $p<0.05$) have been reported. Healing time in the group treated with TMR® was $84.46±54.38$ days versus $148.54±78.96$ days in the control group (statistically significant; $p<0.01$). At the end of a 12-month follow-up period all the patients in the treated group were healed, while in control group three patients were still ulcerated.

### 4. The main mechanisms of magnetic resonance therapy
### 4.1. Cyclotron resonance

In this Section we discuss the physical mechanisms which can be responsible for the wound healing under the exposure to PMF. In the Introduction we have indicated that weak pulsating electromagnetic fields can modify biological processes. Taking into account the complexity of the biological process of wound healing, briefly described in the previous Section, it becomes evident that several physical mechanisms can be responsible for the impact of magnetic therapy. The main facts are the following.

It has been shown that pulsating electromagnetic fields induce cellular transcription (Goodman, 1983). Also it has been clearly demonstrated in (Balcavage, 2015) that EMFs can regulate lymphocyte proliferation *in vitro* and *in vivo*. These effects can take place through the calcium channel activation and modified intracellular Ca levels in the presence of MF.

The basis of several important biological processes, including wound healing, involves ion-selective channels which enable the specific permeation of ions through cell membranes. In particulars, Ca$^{2+}$ ions in mitochondria are directly involved in the metabolism of cells. Voltage-dependent calcium channels are a significant part of the functioning not only of skeletal and smooth muscles, bones (osteoblasts), myocytes, dendrites and dendritic spines of cortical neurons, but also of the brain and peripheral nervous system. Ligand-gated calcium channels are located in plasma membrane. Thus, the functioning of both types of Ca$^{2+}$ channels is reflected on the processes of wound healing. It has been proven experimentally that the intracellular calcium ion concentration increases with exposure to ELF magnetic field, partly due to an increase in the number of activated Ca$^{2+}$ channels. Therefore, one of the possible physical mechanisms can be based on the effects of magnetic fields on functioning of Ca$^{2+}$ ions. In particular, there is a Lorentz force acting on ions as they move through trans-membrane ion channels in the presence of a magnetic field.

Channels are formed by proteins that span the lipid bilayer membrane of the cell. These proteins can change their conformation on application of a mechanical force or an electrical field such that the new conformation can result in an opening or closing of the channel. The magnitude of force required to effect such a change has been estimated to be in the range of 0,2-0,4 pN (Howard, 1989). For some channels, this mechanical force can be actuated by application of an electrical potential difference across the membrane. In this case assuming a membrane thickness about 5nm and a charged centre in the ion channel protein of approximately 2 elementary charges, the necessary force should be 1 pN on the charged centre in the protein. Charged centres within the protein respond to the electric field within the membrane, thus acting as transducers of the force. Such channels are usually described as voltage-gated channels and are typically activated by trans-membrane potential differences in the order of 20mV. In the presence of a magnetic field, Lorentz force, defined in Eq. (1), acts on ions as they move through trans-membrane ion channels. Drift velocity of charges in the channel, $V$, is of the order of $3 \times 10^{-2}$ m/s. According to (Balcavage, 1996), the Lorentz force strong enough to activate nearby voltage gated ion channels, can be achieved in ELF magnetic fields as weak as 100 microT. This can be compared with the calculations of (St Pierre, 2000) according to which the Lorentz force of the magnitude of 1pN requires the magnetic field intensity B =$3 \times 10^8$ microT and the forces generated due to the Lorentz force of the magnetic field on the cation is small as compared with the forces required to activate an ion channel protein conformation change associated with the gating of the channel. These estimates are based on the pure classical 'mechanical' models. Obviously, one can hardly expect the ballistic transport of ions through the trans-membrane channels. There is experimental evidence that ion transport occurs in the form of the travelling waves (Das, 1994). Secondly, the calculations of forces necessary for the conformational changes of proteins, based on classical mechanical models, can give only rough results. According to (Davydov, 1985), a self-consistent account of interaction between the carried charges and elastic degree of freedom of macromolecules (so called electron-lattice interaction) shows that propagation of charges along alpha-helical proteins is significantly facilitated and leads to self-trapping of charges in nonlinear soliton states (this will be discussed below). In an alpha-helical protein soliton propagation is accompanied by a local increase of the protein radius (Davydov, 1978, Brizhik, 2004), which can lead to the effective increase of the ion channel radius (comp. opening of the channel). But even these classical estimates show that weak magnetic fields can facilitate permeability of Ca channels. Indeed, it has been demonstrated experimentally that neuronal ion channels are sensitive to extremely low frequency weak electric field effects (Mathie, 2003).

Another possible mechanism can be related with the effects of MFs on the cardiovascular system (McKay, 2007). These effects on the microcirculation and vasculature can be different depending on the initial state of the cardiovascular system and on medical history of the patient. Thus, it has been reported in (McKay, 2007) that 10 of 27 studies showed vasodilatory effect, increased blood flow or increased blood pressure; 3 studies showed a decrease in blood perfusion/pressure; 4 studies showed no effect; and in 10 studies MF triggered either vasodilation or vasoconstriction depending on the initial tone of the vessels.

Functioning of the cardio-vascular system depends significant on the effectiveness of respiration and accompanying it redox processes (Lehninger, 1972). These processes occur together with the charge transport in the so called Krebbs cycle, which involves the electron transport chain.

Such a chain represents a series of macromolecules onto which electrons can be transferred via redox reactions, so that each compound plays the role of a donor for the 'preceeding' molecule and acceptor for the 'succeding' molecule. Some molecules in the electron transport chains, like quinone or cytochrome *cyt-c*, posess much smaller molecular weight. They are highly soluble and can move relatively easily outside the mitochondrial membrane, carrying electron from a heavy donor to a heavy acceptor. Some other molecules in the electron transport chain, such as NADH-ubiquinone oxidoreductase, flavoproteids, cytochrome c-oxidase *cyt-aa$_3$* and cytochrome *cyt-bc$_1$* complexes are proteins with large molecular weight and thus they are practically fixed in the corresponding membrane. A significant fraction of these proteins is in the alpha-helical conformation, which can support the transport of electrons in the form of electrosolitons (Davydov, 1985; Scott, 1992). Namely these electrosolitons (also called briefly solitons) are sensitive to the constant and pulsating magnetic fields, as it has been shown above.

Three basic physical mechanisms of biological effects of magnetic fields have been identified in (Brizhik, 2015) depending on the hierarchy level of the matter organisation at which such effects takes place: molecular, supramolecular and system mechanisms. Molecular mechanism involves effects of magnetic fields on ions, radicals, paramagnetic particles with electrons and spin, molecules, macromolecules. Supramolecular mechanism involves effects of MF on membranes, mitochondria, microcrystalls, cell nuclei, cells etc. System mechanism is based on the synergetic effect of molecular and supramolecular effects and is manifested on the level of the biological system (endocryne, cardiovascular, nerve, etc.). The biological effects via this latter mechanism are more delayed in time, since they result from the primary effects of the first two mechansims, which are processed with time by the corresponding system and by the whole organism.

Electromagnetic field interaction with living tissues can be realized through these three main mechanisms, affecting energy transfer, charge (matter) transport, and information transport/exchange. Various cellular components, processes, and systems can be affected by EMF exposure. Although there are numerous studies and hypotheses that suggest that the primary site of interaction are membranes, there are also numerous *in vitro* studies indicating that systems, including cell-free systems, can be responsive to EMFs (Goodman, 1995). In particular, extremely low frequency electromagnetic fields can affect cellular responses *in vitro* tests, causing immune cell activation (Simkó, 2004).

Dependence of the biological effects of oscillating magnetic fields on the frequency indicates the resonant character of such effects. Among such resonant mechanisms the first to be mentioned, is the cyclotron resonance. In biological systems there are free electrons, various ions and groups of ions, which are sensitive to the alternating magnetic fields. Due to the Lorentz force, a charge *q=Ze* (an electron, *Z*=1, or an ion of the charge *Ze*) of the mass *m* in a static magnetic field of the intensity *B* moves along the circular trajectory with the angular frequency

$$\omega_c = \frac{Ze}{m} B . \qquad (2)$$

The electromagnetic oscillating field of the frequency *f* resonates with charges which have a mass-to-charge ratio *m*/Z that satisfies the relation

$$\frac{\omega_c}{Z} = \frac{e}{2\pi f} B . \qquad (3)$$

Ion and electron cyclotron resonances are widely known in conventional physical systems. One of the direct observations of this effect in living organisms is the registration of intracellular calcium oscillations in a T-cell line after exposure of the culture to extremely-low-frequency magnetic fields with variable frequencies and flux densities (Lindstrom, 1995). Pulsed EMFs affect the intracellular calcium concentrations in human astrocytoma cells, acting both on intracellular Ca$^{2+}$ stores and on the plasma membrane (Pessina, 2001); 50 Hz electromagnetic fields affect voltage-gated Ca$^{2+}$ channels and their role in modulation of neuroendocrine cell proliferation and death (Grassi, 2004), etc.

Such circular motion of a charge is superimposed with its uniform axial motion, resulting in a <u>helical</u> or a more complex trajectory. The complex trajectory of charge (ions in ion channels, electrons in a charge transport chain, etc) in a highly polar medium of a cell causes excitations of local vibrational modes of biological molecules, which in their turn can result in some changes of their conformations and ability to bind with other molecules.

The ion cyclotron resonance can take place not only in ion channels. Catalysis of DNA transcription involves metal ions (Mg, Zn, ...) and thus, one can expect the effects of magnetic fields on the DNA transcription. Indeed, a significant magnetic isotope effect for magnetic ions $^{199}$Hg, $^{25}$Mg, $^{67}$Zn, $^{43}$Ca takes place in DNA transcription processes (Jasti, 2001; Bawin, 1978) (see also Section 5). Worth mentioning here, that DNA does not appear to be significantly altered directly by low-intensity EMF even at relatively high frequencies, moreover it is the case of ELFs. Nevertheless, MF can affect DNA transcription. It has been suggested by several authors that nonlinear excitations, in particular solitary waves, could play a fundamental functional role in the process of DNA transcription, effecting the opening of the double chain needed for RNA Polymerase to be able to copy the genetic code (Dauxois, 2006). The first step in genome expression is DNA transcription from the original DNA template contained in the cell to a copy – the RNA messenger – which will then be used as a 'master copy' in determining protein sequences in accordance with genetic information. The evolutionary advantage of such a messenger is obvious: in this way, the original DNA is opened – and thus less protected – for a time as small as possible. The transcription process is carried out by a specific enzyme, the RNA Polymerase (RNAP). The dynamical aspect of the transcription is very important and can be briefly described as follows. The RNAP opens a 'transcription bubble' of a size of about 15–20 base pairs, and then travels along the DNA chain keeping the size of the open region relatively constant, i.e., providing at the same time opening of the chain in front of it and closing back the one behind. The RNA Polymerase travels along the DNA chain at a speed of several tens or hundreds base pairs per second. Since each base pair is linked by two or three hydrogen bonds, the energy involved in such a process, even considering only the one to open (and close) the DNA chain, is of the order of at least hundred, if not thousands, hydrogen bonds per second. It has been shown theoretically that there are nonlinear excitations travelling along the DNA chains, causing a local opening of the double chain. The RNAP could then travel along these chains and read the DNA sequence without the need to have the energy necessary to open the whole double chain. Another advantage of this theory is connected with the closing of the double chain after the RNAP has passed: if this is done in a non-coordinated way, it will generate a substantial quantity of random motion and thus of thermal energy; if instead this corresponds to the traveling of a nonlinear excitation, practically no thermal energy is generated when the system recovers its local fundamental state.

Magnetic field can also modify the population distribution of the nuclear and electronic spin states and cause alignment of the magnetic moments of the components of the radical pairs, which can affect concentration of free radicals, as it has been suggested in (Barnes, 2015). These changes can modify biological processes.

### 4.2. Molecular solitons and electrosolitons

Another sensor of the external magnetic field can be molecular solitons and electrosolitons, which we call below by the same term 'soliton' (Davydov, 1985; Scott, 1992). We remind here that a molecular soliton (called also as Davydov's soliton) is a nonlinear bound state of a molecular excitation like AMID-I excitation and local deformation of quasi-one-dimensional molecular chain, like, for instance, polypeptide chain in macromolecules. Similarly, an electrosoliton is a bound state of an electron and local deformation of the macromolecule. It has been shown theoretically that Davydov's solitons provide storage and transfer of the ATP hydrolysis energy in cells (Davydov, 1985; Scott, 1992).

According to (Brizhik, 1984), electrosolitons provide transport of electrons in certain steps of redox processes during respiration, as is summarized below. This transport takes place along the so-called electron transport chain (Lehninger, 1972; Helms, 2008). Such chain represents a series of biological molecules that transfer electrons from one to another via redox reactions, so that each compound plays the role of a donor for a molecule 'on the left' and acceptor for the molecule 'on the right'. In eukaryotes the electron transport chain is located in inner mitochondrial membrane, where oxidative phosphorylation with ATP synthase takes place (Lehninger, 1972; Helms, 2008). Some molecules in the electron transport chains, like quinone or cytochrome *cyt-c*, have relatively small molecular mass. They are highly soluble and can move relatively easy outside the mitochondrial membrane, carrying an electron from a heavy donor to a heavy acceptor. In theoretical studies such electron transport systems are modeled as complexes which include a donor molecule weakly bound to a bridge molecule, which in its turn is weakly bound to an acceptor molecule. The bridge itself can be modeled as some potential barrier through which the electron tunneling takes place (see, e.g., (Helms, 2008) and references therein. In some other studies the bridge is modeled as a molecule with super-exchange electron interaction taken into account. It has been shown that properties of electrons in these systems differ little from properties of free electrons. Therefore, their cyclotron resonance frequency is close to the frequency determined in Eq. (2) setting $Z=1$.

Some other molecules in the electron transport chain, such as NADH-ubiquinone oxidoreductase, flavoproteids, cytochrome c-oxidase *cyt-aa$_3$* and cytochrome *cyt-bc$_1$* complex are proteins with large molecular weight, and, thus, they are practically fixed in the corresponding membrane (Lehninger, 1972). Qualitatively and quantitatively different situation takes place for electrons, when they are transported through these proteins of large molecular mass. First of all, a significant part of such proteins is in alpha-helical conformation, which is stabilized by relatively weak hydrogen bonds between every fourth peptide group (a group of atoms H-N-C=O), so that along the helix there are three hydrogen-bounded polypeptide chains. The softness of hydrogen bonds and quasi-one-dimensional structure of polypeptide chains suggests a possible significant role of the electron-lattice interaction in them. Indeed, it has been shown (Davydov, 1985) that this electron-lattice interaction is relatively strong and that it results in a self-trapping of electrons: electrons, transferred into a protein from a donor molecule, create a local deformation of the protein. Such deformation acts as a potential well, which attracts an electron. As a result, a bound state of an electron and lattice deformation is formed. This state is described by the system of coupled nonlinear equations for the electron wave-function, $\Psi(x,t)$, and lattice deformation, $\rho(x,t)$:

$$i\hbar\frac{\partial\Psi(x,t)}{\partial t}+J\frac{\partial^2\Psi(x,t)}{\partial x^2}+\chi\rho(x,t)\Psi(x,t)=0, \quad (4)$$

$$\frac{\partial^2\rho(x,t)}{\partial t^2}-V_{ac}^2\frac{\partial^2\rho(x,t)}{\partial x^2}+\frac{\chi}{M}\frac{\partial^2}{\partial x^2}|\Psi(x,t)|^2=0 \quad (5)$$

Here $x$ is the coordinate along the polypeptide chain, $J$ is the electron exchange interaction constant coming from the overlap of electron wave-functions on the neighboring peptide groups of mass $M$, $\chi$ is the electron-lattice coupling constant, $V_{ac}$ is the velocity of sound in a polypeptide chain. In other words, the first equation describes an electron in the potential, created by the lattice deformation. This deformation is proportional to the probability of the electron presence at the given site, as it follows from the solution of Eq. (5):

$$\rho(x,t)=\frac{\chi}{w(1-s^2)}|\Psi(x,t)^2|. \quad (6)$$

Here $w$ is the elasticity of the hydrogen bond, $s$ is the velocity of an electron measured in units of the velocity of sound, $s=|V|/V_{ac}$. It is worth to recall that the electron wave function is normalized to 1, since square of its modulus determines electron probability.

Substituting this solution into Eq. (4), we derive the so-called nonlinear Schroedinger equation

$$i\hbar\frac{\partial\Psi(x,t)}{\partial t}+J\frac{\partial^2\Psi(x,t)}{\partial x^2}+2Jg|\Psi(x,t)|^2\Psi(x,t)=0, \quad (7)$$

where $g$ is the dimensionless nonlinearity constant

$$g=\frac{\chi^2}{2Jw(1-s^2)}\equiv\frac{g_0}{(1-s^2)}, \qquad g_0=\frac{\chi^2}{2Jw}. \quad (5)$$

Equation (7) at s<1 admits the so-called soliton solution:

$$\Psi(x,t)=\Psi_s(x,t)\equiv\frac{1}{2}\sqrt{g}\,Sech\left[g(x-x_0-Vt/a)/2\right]\exp\left(im_eVx/\hbar+i\varphi_s(t)\right). \quad (8)$$

It follows from the solution (8), that in a soliton state the electron probability, $P(x,t)=|\Psi_s(x,t)|^2$, and, according to the relation (6), the chain deformation, are localized in space with the width of the localization

$$l_s = \frac{\pi a}{g}, \qquad (9)$$

as it is shown in Fig.5. Here $a$ is the distance between the neighboring peptide groups, $x_0$ is the position of the soliton center of mass at the initial time moment $t=0$, and $\varphi_s$ is the time-depending phase of the soliton. Such soliton corresponds to the ground electron state (the state with the lowest energy) at intermediate values of the electron-lattice coupling and small values of the nonadiabaticity parameter. Namely such conditions are fulfilled in polypeptides (Davydov, 1985), Scott, 1992).

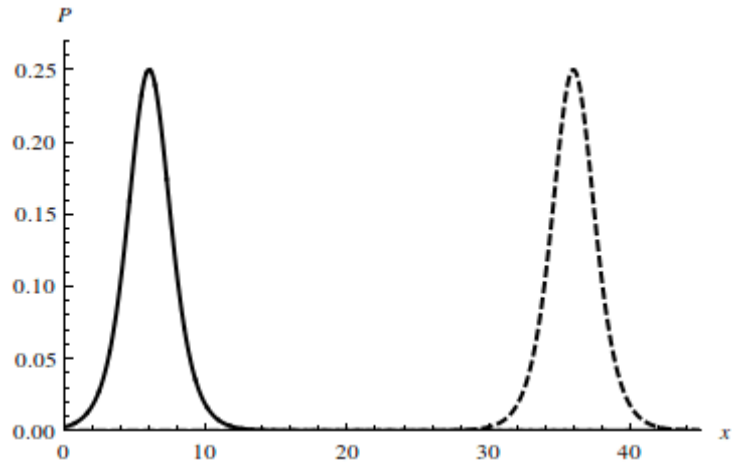

Fig. 5. Probability of electron localization, $P$, in a soliton state along the chain axis $x$ at the parameter values $g=1$, $x_0 = 5$ at time moments $t=0$ (solid line) and $t = 30a/V$ (dashed line).

This mathematical model describes also self-trapping of the AMID-I excitation in alpha-helical proteins (Davydov, 1985). Such excitation is formed in the result of the hydrolysis of the ATP (adenosine triphosphate) molecule into ADP (adenosine diphosphate) molecule. Namely soliton mechanism is the only one which explains exceptional effectiveness of ATP hydrolysis energy storage and transfer in cells (Scott, 1992).

From Eqs. (5) and (7) it follows also, that the electron and lattice deformation propagate along the polypeptide chain with constant velocity $V$ from one end to the other one, from a donor to an acceptor molecule, as a coherent localized wave (see Fig. 3 and Eq. 6). Such a wave does not emit phonons because it moves with the velocity less than the velocity of sound, $|V|<V_{ac}$, and is exceptionally stable, able to propagate on macroscopic distances in view of the extremely low dissipation of energy and the nonlinear nature of it formation.

In view of the binding of an electron with the lattice deformation, the effective mass of a soliton is bigger than the mass of a free electron, $m_e$:

$$m_s = m_e(1+\delta), \quad \delta = \frac{Jg_0^2}{3mV_{ac}^2}. \qquad (10)$$

Therefore, the cyclotron resonance frequency of an electron in the soliton state is different from the frequency of a free electron, and in the first approximation can be estimated as:

$$\omega_{c,s} = \frac{e}{m_s}B. \qquad (11)$$

A more rigorous study of the dynamics of an electrosoliton in a magnetic field requires extension of the one-dimensional model to the 3-dimensional space, when all functions depend on three spatial coordinates:

$$\Psi = \Psi(x,y,z;t), \qquad \rho = \rho(x,y,z;t). \qquad (12)$$

The state of an electron in a strongly anisotropic crystal, like polypeptide macromolecule, in the absence of the magnetic field is described by the Shroedinger equation

$$i\hbar \frac{\partial \Psi(x,y,z;t)}{\partial t} = H\Psi(x,y,z;t), \qquad (13)$$

with the Hamiltonian function

$$H = \sum_{\nu=1,2,3} \frac{1}{2m_\nu} p_\nu^2 - \chi\rho(x,y,z;t). \qquad (14)$$

Here $p_\nu = -i\hbar \partial/\partial x_\nu$ is the corresponding component of the momentum operator of an electron, and $1/m_\nu$ is the tensor of the electron effective mass.

Deformation of the polypeptide chain oriented along $x$-direction, is determined by the equation which is the generalization of Eq. (5) to a 3D space:

$$\frac{\partial^2 \rho(x,t)}{\partial t^2} - V_{ac}^2 \frac{\partial^2 \rho(x,t)}{\partial x^2} + \frac{\chi}{M} \frac{\partial^2}{\partial x^2} \int dy dz |\Psi(x,y,z;t)|^2 = 0 \qquad (15)$$

The solution of the system of equations (13), (15) is a wave function, which is the product of plane waves in perpendicular to the chain direction, and soliton wave function in the chain direction:

$$\Psi(x,y,z;t) = \frac{1}{\sqrt{l_y l_z}} \Psi_s(x - x_0 - Vt/a) \exp(i\vec{k}\vec{r} - iEt/\hbar). \qquad (16)$$

Here $l_y, l_z$ are the sizes of the crystal in the transverse to the chain directions, $\hbar k_x = m_x V$, and $\Psi_s$ is defined in Eq. (8).

In the presence of the external magnetic field $\vec{B}$ dynamics of an electron is governed by the system of equations (13),(15), in which the Hamiltonian includes the vector-potential of the magnetic field, $\vec{A} = rot\vec{B}$,

$$H = \sum_{\nu=1,2,3} \frac{1}{2m_\nu} \left( p_\nu - \frac{e}{c} A_\nu \right)^2 - \chi\rho(x,y,z;t). \qquad (17)$$

Below we consider two cases, when the magnetic field is parallel or perpendicular to the chain. Any other orientation of the field can be reduced to the vector combination of these two cases.

### 4.3. `Electrosoliton in a longitudinal magnetic field

In the case of the field, parallel to the chain, we can set $\vec{B} = (B,0,0)$, and choose the vector-potential in the form $\vec{A} = (0,-Bz,0)$. The Hamiltonian (17) can be represented in the form of the sum of the soliton part, $H_s$, and transversal part, $H_t$:

$$H_s = -\frac{\hbar^2}{2m_x} \frac{\partial^2}{\partial x^2} - \chi\rho(x,t), \qquad (18)$$

$$H_t = -\frac{1}{2m_y} \left( -i\hbar \frac{\partial}{\partial y} + \frac{e}{c} Bz \right)^2 - \frac{\hbar^2}{2m_z} \frac{\partial^2}{\partial z^2}. \qquad (19)$$

Therefore, the electron wave function can be represented in the form of the product

$$\Psi(\vec{r};t) = \Psi_s(x,t)\Psi_{tr}(y,z;t), \qquad (20)$$

where the soliton wave function is defined by Eq. (8), and the transversal wave function is given by the relation

$$\Psi_{tr}(y,z;t) = \frac{1}{\sqrt{l_y}} \varphi(z) \exp(ik_y y - iE_t t/\hbar),$$

where $\varphi(z)$ satisfies the following equation:

$$\left[ -\frac{\hbar^2}{2m_z} \frac{\partial^2}{\partial z^2} + \frac{e^2 B^2}{2m_y c^2} (z - z_0)^2 \right] \varphi(z) = E_t \varphi(z). \qquad (21)$$

Here
$$z_0 = \frac{|B|e}{m_B c^2}. \qquad (22)$$

It follows from Eq. (21) (see (Brizhik, 1990)) that function $\varphi(z)$ is the function of the harmonic oscillator, $\varphi(z) = \varphi_n(z - z_0)$, with the energy

$$E_{tr} = \hbar\omega_B\left(n + \frac{1}{2}\right), \qquad n=0,1,2,\ldots, \qquad (23)$$

which is determined by the cyclotron frequency

$$\omega_B = \frac{|B|e}{m_B c}, \qquad m_B = \sqrt{m_y m_z}. \qquad (24)$$

The total energy of the electron with account of the interaction of its spin S=1/2 with the magnetic field, is given by the relation

$$E = -\frac{1}{12}Jg_0^2 + \frac{1}{2}m_s V^2 + \hbar\omega_B\left(n + \frac{1}{2}\right) + 2\mu_B S_x B, \qquad (25)$$

where $\mu_B = e\hbar/2m_e c$ is Bohr magneton and $S_x = \pm 1/2$ is the projection of the electron spin on the direction of the magnetic field.

The energy level (23) is degenerate, the degree of degeneracy is determined by the number of the possible values $k_y$ at which the equilibrium position of the electron $z_0$ is located inside the macromolecule.

### 4.4. Electrosoliton in a transverse magnetic field

Let us now consider the case when the magnetic field is perpendicular to the chain, for instance, is oriented along z-axis $\vec{B} = (0,0,B)$. In this case we can choose the vector-potential in the form $\vec{A} = (0, Bx, 0)$.

Then the Hamiltonian (17) can be represented in the form

$$H = -\frac{\hbar^2}{2m_x}\frac{\partial^2}{\partial x^2} + \frac{1}{2m_y}\left(-i\hbar\frac{\partial}{\partial y} - \frac{e}{c}Bx\right)^2 - \frac{\hbar^2}{2m_z}\frac{\partial^2}{\partial z^2} - \chi\rho(x,t). \qquad (25)$$

This Hamiltonian leads to the system of equations which includes the Scroedinger equation

$$i\frac{\partial\Psi(\tilde{x},\tau)}{\partial\tau} + \frac{1}{2}\frac{\partial^2\Psi(\tilde{x},\tau)}{\partial\tilde{x}^2} + \frac{\chi}{2Jg}\rho(\tilde{x},\tau)\Psi(\tilde{x},\tau) = \varepsilon(\tilde{x} - \tilde{x}_0)^2 \Psi(\tilde{x},\tau) \qquad (26)$$

and the corresponding equation for the chain deformation (5). Here the dimensionless units are introduced:

$$\tilde{x} = \sqrt{g}x, \qquad \tau = \frac{2Jgt}{\hbar}, \qquad (27)$$

Here parameter $g$ is determined in Eq. (5) and the coefficient is defined

$$\varepsilon = \frac{m_x \omega_B^2 a^2}{4Jg^2} = \frac{1}{2g^2}\frac{m_x}{m_y}\varepsilon', \qquad \varepsilon' = \frac{e^2 B^2 a^4}{\hbar^2 c^2}. \qquad (28)$$

It is a small parameter even for strong magnetic fields. Hence, we can search the solution for the chain deformation in the form:

$$\rho(\tilde{x},t) = \rho_0(u) + \varepsilon\rho_1(u), \qquad u = x - \xi(t). \qquad (29)$$

Substituting anzats (29) into Eq. (5), we see that

$$\rho_0(u) = \frac{\chi}{w(1-s^2)}|\Psi(u)|^2, \qquad V = a\frac{d\xi(t)}{dt}, \qquad (30)$$

and obtain the equation for $\rho_1(t)$:

$$\frac{d\rho_1(u)}{du} = -\frac{a^2}{\varepsilon V_{ac}^2(1-s^2)}\frac{d^2\xi(t)}{dt^2}\rho_0(u). \qquad (31)$$

Substituting Eqs. (29)-(31) into Eq. (27), we see that it can be reduced to the nonlinear Scroedinger equation with the right hand side

$$i\frac{\partial \Psi(\tilde{x},\tau)}{\partial \tau} + \frac{1}{2}\frac{\partial^2 \Psi(\tilde{x},\tau)}{\partial \tilde{x}^2} + |\Psi(\tilde{x},\tau)|^2 \Psi(\tilde{x},\tau) = i\varepsilon R[\Psi(\tilde{x},\tau)] \quad (32)$$

where

$$R[\Psi] = -i\left[(\tilde{x}-\tilde{x}_0)^2 - \frac{\chi}{2Jg}\rho_1(\tilde{x})\right]\Psi \quad (33)$$

Eq. (32) with the right hand side (33) has been studied in (Brizhik, 1990) in frame of the nonlinear perturbation theory based on the inverse scattering problem for the Nonlinear Schroedinger equation. We will skip here the corresponding calculations and represent the final results. Thus, the soliton wave function is given by the expression

$$\Psi_s(x,t) = 2\nu Sech(\varsigma)e^{i\frac{\mu}{\nu}\varsigma + i\eta}, \quad \varsigma = 2\nu(x-\xi) \quad (34)$$

where $\mu$ and $\nu$ are the parameters which determine the width and amplitude of the soliton and, and $\xi$ is the centre of mass coordinate of the soliton. All these parameters weakly depend on time:

$$\mu(t) = \mu_0 \cos\left(\sqrt{\frac{2\varepsilon}{1+\delta}}\frac{2Jg}{\hbar}t\right), \quad \nu(t) = \frac{1}{4}\sqrt{g} = const \quad (35)$$

$$\xi(t) = 2\mu_o\sqrt{\frac{1+\delta}{2\varepsilon}}\sin\left(\sqrt{\frac{2\varepsilon}{1+\delta}}\frac{2Jg}{\hbar}t\right) + x_0, \quad (36)$$

where $\delta$ is defined in Eq. (10) and

$$\mu_0 = \frac{m_x Va}{2\hbar\sqrt{g}}. \quad (37)$$

Substituting (35)-(37) into the expression (34), we rewrite it as

$$\Psi_s(x,t) = \frac{1}{2}\sqrt{g}Sech\left[g\left(x - x_0 - \frac{V}{\omega}\sin(\omega t)\right)/2a\right]\exp(i\phi(x,t)), \quad (38)$$

Here

$$\omega = \frac{\omega_B}{\sqrt{1+\delta}} \quad (39)$$

and $\omega_B$ is determined by the expression (24).

Therefore, soliton wave function in the transverse magnetic field is given by the expression (38) and describes propagation of the localized wave package with the envelope which coincides with the envelope of a free soliton, but whose velocity of propagation is oscillating in time with frequency of oscillations which depends on the magnetic field:

$$V(t) = V\cos(\omega t) \quad (40)$$

The cyclotron frequency of the soliton is determined by its effective mass (10), which is bigger than the cyclotron mass of a free electron. This means that although the magnetic field acts directly on the electron charge, it also acts on the macromolecule due to binding of the electron in the soliton state with the self-induced deformation of the macromolecule. The oscillating character of the propagation of soliton and bisoliton with frequency (40), respectively, according to the relation (30), is accompanied by the propagation of the local deformation of the polypeptide chain, $\rho(x,t)$, which also is the oscillating function of time. This deformation will excite additional vibrational modes in the polypeptide chain and can change its conformation, which can be reflected on the conformation-function relation. Such structural changes can be critical for many processes, from cellular communication through membrane ion channels to oxygen uptake and delivery by hemoglobin. Indeed, the long-range protein vibrational modes have been optically registered in (Acbas, 2014). In particular, binding of the so-called T cells which are known to be the key modulators of inflammation, can be affected by the MF. It has been shown that 0.1 mT, 60 Hz EMFs induce a 20% mean-increase in anti-CD3 binding to T cell receptors (TcRs) of Jurkat cells, a T lymphocyte cell line (Balcavage, 2015). It also has been shown there that 60 Hz sinusoidal EMFs and a commercial bone healing EMF modulate signal transduction pathways that regulate lymphocyte proliferation and that are normally triggered by activating the Jurkat TcR. Similar EMF effects have been shown in human peripheral blood lymphocytes (hPBLs), exposed to EMFs in culture and in rat PBLs, when donor animals are exposed to a bone healing field (21 days, 4 hr/day).

Thus, we see that the magnetic field affects the electrosoliton transport, and, therefore, it can affect the redox processes. Indeed, the electromagnetic induction of protection against oxidative stress has been demonstrated experimentally (Di Carlo, 2001).

## 5. Experimental support of the described above mechanisms

Below we describe the data, which in our opinion provide the experimental support of the mechanisms developed in the previous Section. They are based on the clinical experience (study on surgical patients) (Abbruzzese, 2015) which shows statistically significant evidence of healing process acceleration.

*In vitro* tests both on healthy and diabetic tissue under the exposure to TMR® show not only an acceleration of fibroblast and endothelial cells growth after the exposure, but also the reduction of oxidative stress condition both in fibroblast and endothelial cells. Clinical experience of TMR® usage in a double-blind exploratory clinical study results in a dramatic reduction of lesion surface (as compared to conventional treatment) and lesion volume.

This data shows the direct correlation between the decrease of the oxidative stress and wound healing. The oxidative stress, in its turn, results from the inhibition of the redox processes. As it follows from our theoretical study in the previous Section, the magnetic field can enhance/restore charge transport and, thus, improve the redox processes, reducing oxidative stress.

In the previous Section we have indicated, that soliton mechanism describes the storage and transfer of the energy of ATP hydrolysis and have shown sensitivity of solitons to MF. Indeed, it has been shown experimentally that ATP synthesis can be induced by PEF (Carpenter, 1994) and that MF affects enzymatic ATP synthesis (Buchachenko, 2008).

In inflammation massive infiltration of T-lymphocytes, neutrophils and macrophages into the damaged tissue takes place (Moller, 2000). Based on the presence of $A_{2A}$ adenosine receptors in human neutrophils, it has been suggested that adenosine plays an important role in modulating immune and inflammatory processes. Therefore, activation of $A_{2A}$ receptors by PEMFs may have a relevant therapeutic effect (Zhang, 1997; Zhang, 2002). Neutrophils are the most abundant white cells in the peripheral blood and are usually the first cells to arrive at an injured or infected site. Adenosine, interacting with specific receptors on the surface of neutrophils, is an endogenous anti-inflammatory agent (Andrew,1993). The activation of $A_{2A}$ receptors in human neutrophils affects the immune response in auto-immune and neurodegenerative diseases and decreases inflammatory reactions (Yasuda, 1954). Exposure to PEMFs can suppress the extravascular oedema during early inflammation (Lee, 1997). It has also been demonstrated that the complete healing of wounds depends on the presence of $A_{2A}$ adenosine receptor agonists (Montesinos, 1997). It has been reported that PEMFs mediate positive effects on a wound healing, controlling the proliferation of inflammatory lymphocytes and resulting in beneficial affects on inflammatory disease (Regling, 2002). Exposure to PEMF can trigger a more complex biologic response such as cell proliferation as it is evident from some clinical results (Corbellini, 2006; Yan, 2010; Gaetani, 2009). It can induce an increase in the proliferation of human articular chondrocytes suggesting an important role also in cartilagine repair (Ishido, 2001).

In biological systems there is a whole set of signaling mechanisms, known as transduction pathways, which involve various biochemical reactions, some of which take place with participation of ionic radicals (Lehninger, 1972). Cells as no other known system can convert one type of stimulus into another, using chains of biochemical reactions involving enzymes. Enzymes are activated by specific molecules, called messengers, for instance, the cellular calcium ion, $Ca^{2+}$. Therefore, they can be affected by oscillating magnetic fields. The experimental evidence of this has been given in Section 2 (see also (Brizhik, 2015)).

It has been reported in (Pazur, 2007) that through the ion cyclotron resonances the magnetic field can be useful in the regenerative medicine. Oscillating magnetic field effects human epithelial cell differentiation through the ion cyclotron resonance (Gao, 2005; Buchahchenko, 2014). According to A. Liboff, calcium and potassium ions can be specifically activated by the magnetic field through this effect, which enhances their transport through membrane ion channels, thereby altering signaling mechanisms and cellular function (Cadossi, 1992). These signals are mediated in cells by the cytoplasm, in which water is one of the main components

Another important biological process, sensitive to magnetic field, is DNA transcription by polymerase. Like ATP synthesis, catalysis of DNA transcription involves metal ions (Mg, Zn, ...) and also shows significant magnetic isotope effect for magnetic ions $^{199}$Hg, $^{25}$Mg, $^{67}$Zn, $^{43}$Ca (Jasti, 2001; Bawin, 19878). The transcription is initiated by the mechanical unwinding of DNA at the promoter site. The RNA polymerase (RNAP) can be viewed as a torque wrench, which opens a 'transcription bubble' of a size of about 15–20 base pairs, that travels along the DNA as a solitary wave keeping the size of the open region almost constant, opening the double chain in front of the RNAP and closing the chain behind it. It is assumed that the transcription bubble travels along the chain in the shape of the soliton used by RNAP to access the base sequence (Dauxois, 2006). This process can be altered by pathological conditions. The TMR™ can help functioning of normal cells reactivating the physiological processes. In particular, as we have discussed above, the interaction between the weak magnetic field of the TMR™ with biological tissue, can be related to the mechanical action produced by the field itself on the cells in the form of the Lorentz force Eq. (1). Also a rotational torque acts on DNA topology, helping launching soliton, and so stimulating cell replication. Indeed, it has been shown experimentally that pulsating EMFs can induce cellular transcription (Goodman, 1983). The hypothesis that transcription responses depend on pulse characteristics was evaluated by using two pulses in clinical use, the repetitive single pulse and the repetitive pulse train. These pulses produced different results from each other and from controls. The single pulse increased the specific activity of messenger RNA after 15 and 45 minutes of exposure.

The mechanism of magnetic isotope effect is connected with the interaction of the unpaired electron of cation-radical $Mg^+$ with the nuclei, which results in the change of the electron spin of the pair: spin conversion of the pair from a singlet to a triplet state takes place. Such change of the spin opens a new reaction channel in a radical pair. It has been shown that enzymes with magnetic ions $^{25}Mg^{2+}$ can be activated by magnetic field, while enzymes with a nonmagnetic $^{24}Mg^{2+}$ are inhibited (Foletti, 2010). In a similar way external magnetic field can control such reactivity.

The soliton mechanism of the redox processes in respiration, described briefly in Section 4, is supported by some experimental data as well. In particular, it has been shown that weak magnetic RF and SMFs increase rate of hemoglobin deoxygenation in a cell (Jasti, 2001). Catalysis of oxidation of nicotinamide adenine dinucleotide (NADH), which participates in the redox processes, by molecular oxygen, is performed by peroxidase enzyme. This oxidation is an oscillating reaction with the period of oscillations approximately 100 s, during which concentrations of NADH and $O_2$ oscillate. It has been shown that these reactions depend on the period of oscillations and intensity of the magnetic field in the interval 1000-4000 Gs (Huang, 1997).

It is well known that free radicals are constantly formed in the body during normal metabolic processes and are directly involved in inflammatory processes. When their formation is significantly increased, or protective mechanisms are compromised, a state of oxidative stress will result. If oxidative stress is persistent, it will lead to molecular damage and tissue injury. The antioxidative effect of the MF via the soliton mechanism, helps to reduce free radicals and consequently the oxidative damage, also increasing the antioxidative defense. Oxidative stress and apoptosis in relation to exposure to MF has been discussed above (see Fig.1) and also demonstrated experimentally in (Emre, 2011). This can be one of the mechanisms responsible for the TMR™ therapy as well.

The treatment with specific frequencies of electromagnetic waves corresponding to some optimal regime to optimize the redox balance (rH$_2$) and the acidity (pH) of body fluids to restore the cellular metabolism, has been reported in (Foletti, 2009). A significant improvement in the wound closure and bone fractures healing process, improvement of osteogenesis in osteoporosis have also been shown in (Foletti, 2009; Caneá, 1973; Satter, 1999), as well as in wound healing (De Mattei, 1999).

A significant decrease of ROS production, when the cultures of fibroblasts and endothelial cells were treated with TMR™, has been revealed experimentally (see Fig. 1). The time-dependent decrease of ROS productions, and, hence, of metabolic activity in the cells treated with the PEMFs, is shown in Fig. 6. ROS measurements (Ferroni, 2015) were performed using the OxiSelect™ ROS Assay Kit (measurement of fluorescence level). In the control cells (cells not treated with PEMFs), an increase in ROS production takes place due to the in vitro ageing of the cells.

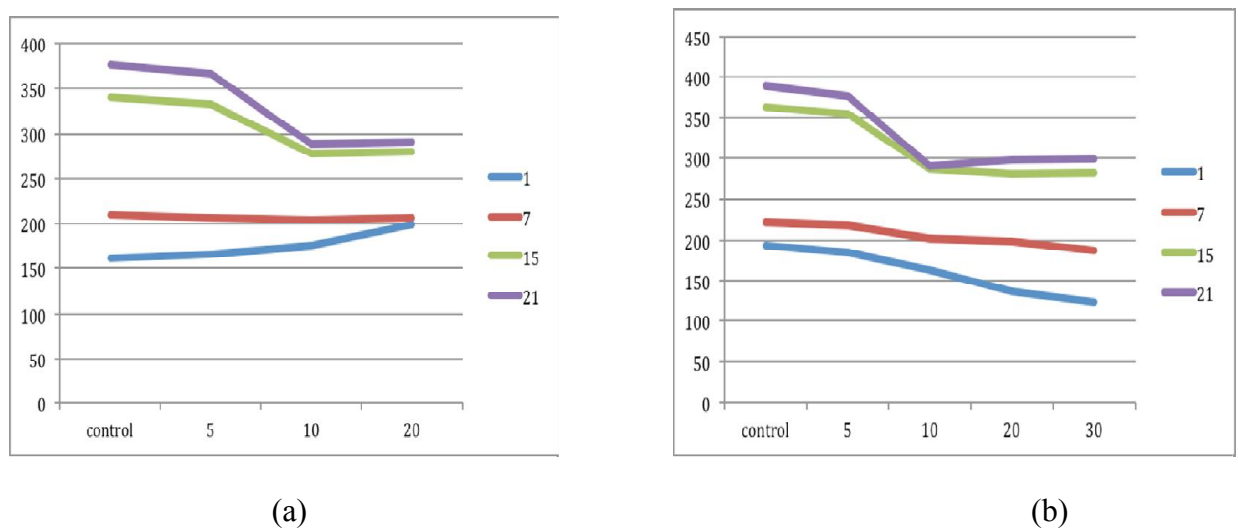

(a)                      (b)

Fig.6. ROS (rfu/s, relative fluorescence units/seconds) as function of time for different duration of treatment with PEMF in (a) fibroblasts and (b) endothelial cells.

These data confirm the ability of TMR™ to influence mitochondrial function, in particular, and the cellular energetics in the whole. These facts indirectly support our model as well. We remind here that fibroblasts are cells that synthesize the extracellular matrix and collagen and play a critical role in wound healing. Endothelial cells are thin flattened cells which line the inside surfaces of body cavities, blood and lymph vessels , and, thus, are also involved in wound healing.

We can't skip the subject of possible contradictions between the data obtained by various authors on the therapeutic effects of MFs and/or low reproducibility of such data. It is worth to recall, that living organisms possess complex structure and organization, which are determined by a huge number of factors (parameters). Due to this, biological systems can be very sensitive to some of these parameters and much less sensitive to others, which makes it difficult to reach a high reproducibility of the experiments with living organisms. Moreover, the biological impact of the external stimuli in general, and of magnetic field in particular, can depend on the phase of the development of biological cells and their synchronization if any, 'history' of the system under study, etc. We have discussed this problem in the aspect of biological effects in our previous paper (Brizhik, 2015). This is the case even at the lower level of living matter organization: the age and state of the cell can profoundly affect the EMF biological response (Goodman, 1995). The effects of magnetic field exposure on microcirculation and microvasculature have been studied in (McKay, 2007) and shown to be different: approximately half of the cited there studies indicated a vasodilatory effect of MFs; the remaining half indicated that MFs could trigger either vasodilation or vasoconstriction depending on initial vessel tone. Indeed, perception and sensitivity to exposure not only depend on field parameters, including the field intensity, frequency, modulation, duration of the exposition, etc., but also can vary for different people, as it follows from numerous studies (see, e.g., (Habash,2007)). In humans, the biological response is dependent not only on exposure at particular magnetic field strengths and frequency, but also on specific shape of magnetic field pulses.

Thus, the results on the use of magnetostimulation in treatment of patients with painful diabetic polyneuropathy reported in (Cieslar, 1995; Weintraub, 2003; Wróbel, 2008), are to a certain extent contradictory. While the positive effects of such treatments in reducing pain intensity, improving quality of life, and decreasing sleep disturbances, etc., have been reported (Cieslar, 1995; Weintraub, 2003). The authors of (Wróbel, 2008) concluded, that the genuine magnetic field exposure had no advantage over sham exposure. Comparison of these results is difficult because of differences between the magnetic field parameters exposure duration, total exposure times and devices used to generate the magnetic fields. The biological response occurs only within a specific amplitude and frequency range, being moderate or absent outside of this 'window'. This can explain the contradicting findings in the studies using different exposure profiles by different authors.

## 6. Conclusions

We conclude that there are several physical mechanisms which cause biological effects of magnetic fields, in general, and, in particular, can result in the positive impact of magnetic therapies, such as, for instance, magnetic resonance therapy, registered as TMR$^®$ therapy, when patients are exposed to low-intensity pulsating electro-magnetic fields at specific patented shapes and low frequencies of pulses. We have reported above "in vivo" confirmation of TMR$^®$ efficacy in acceleration of skin tissue regeneration in wound healing process of diabetic and non-diabetic patients, and, in particular in wound healing of the diabetic foot disease. To understand the working principle of this therapy, we analyzed relevant to it biological effects produced by magnetic fields. Based on these data, we have shown that there is a hierarchy of the possible physical mechanisms, which can produce such effects. These mechanisms act on different levels of the hierarchy of the organization of living organisms and at different time scales.

The mutual interplay between these mechanisms can lead to a synergetic outcome delayed in time, which can affect the physiological state of the organism. We have shown above that soliton mediated energy and charge transport during metabolism is sensitive to magnetic fields, so that such fields can facilitate energy storage and transfer, and enhance redox processes, which, in its turn, can stimulate the healing effect of the organism in general. We have shown that within the soliton mechanism of energy and charge transport, the magnetic field can cause a hierarchy of changes from the primary effect on the dynamics of solitons, to the changes of the conformational states of macromolecules, to the effects on the rate of respiration, and, finally, to the effect on the whole metabolism of the system.

These effects can be summarized as the three main mechanisms which can explain the interaction between Therapeutic Magnetic Resonance TMR™ and biological tissue in wound healing process:

1. Effect of magnetic field on nonlinear excitations involved in the DNA transcription process (RNA Polymerase). This effect is attained in the result of the *local therapy.*

2. Increase of the effectiveness of the ATP hydrolysis energy storage and transfer in cells in the form of Davydov's solitons. This effect is attained in the result of the *local therapy and total body exposure.*

3. Increase of the effectiveness and synchronization of redox processes in the result of effects of magnetic field on electrosolitons. This effect is attained in the result of the both *local therapy and total body exposure.*

It is worth to stress here, that these mechanisms are essentially non-thermal and can be caused by low-intensity MFs, which in fact are used in magnetic therapies. The energetical and dynamical stability of nonlinear soliton states is due to the compensation between the dispersive effects of wave processes and nonlinearity of the system, i.e., electron-lattice interaction. In the result of this interplay solitons propagate along macromolecules with very low energy dissipation, which explains high efficiency of the bioenergetics and DNA transcription. If the closure of the double chain after the RNA polymerase has passed, is done in a non-coordinated way, it will generate a substantial quantity of random motion and, thus, of thermal energy. If, instead, this happens after the passing of a nonlinear excitation, very little thermal energy is generated when the system recovers its ground state.

We conclude that according to biology-histopathology evidence and exploratory clinical investigation, TMR® therapy looks as a promising technique to treat early stage and advanced stage diabetic foot ulcers and significantly accelerates their healing process. The mechanisms and the biological impact of magnetic fields on the processes involved in wound healing, has been summarized as the working principle which has been called as 'Brizhik-Fermi working principle of the TMR® therapy' at the THERESON Scientific Meeting (Piacenza, Italy, June 06, 2014) and is shown schematically in Fig. 7. The knowledge of this working principle can provide information useful for optimizing patient treatment protocols and procedures, and can be used as the guidelines for a further, pivotal clinical investigations related to therapies based on usage of magnetic fields.

**Acknowledgment** The authors acknowledge stimulating discussions with C. Simmi and D. Zanotti from THERESON Company (Italy). This research was carried under the partial support from the Fundamental Research grant of the National Academy of Sciences of Ukraine.

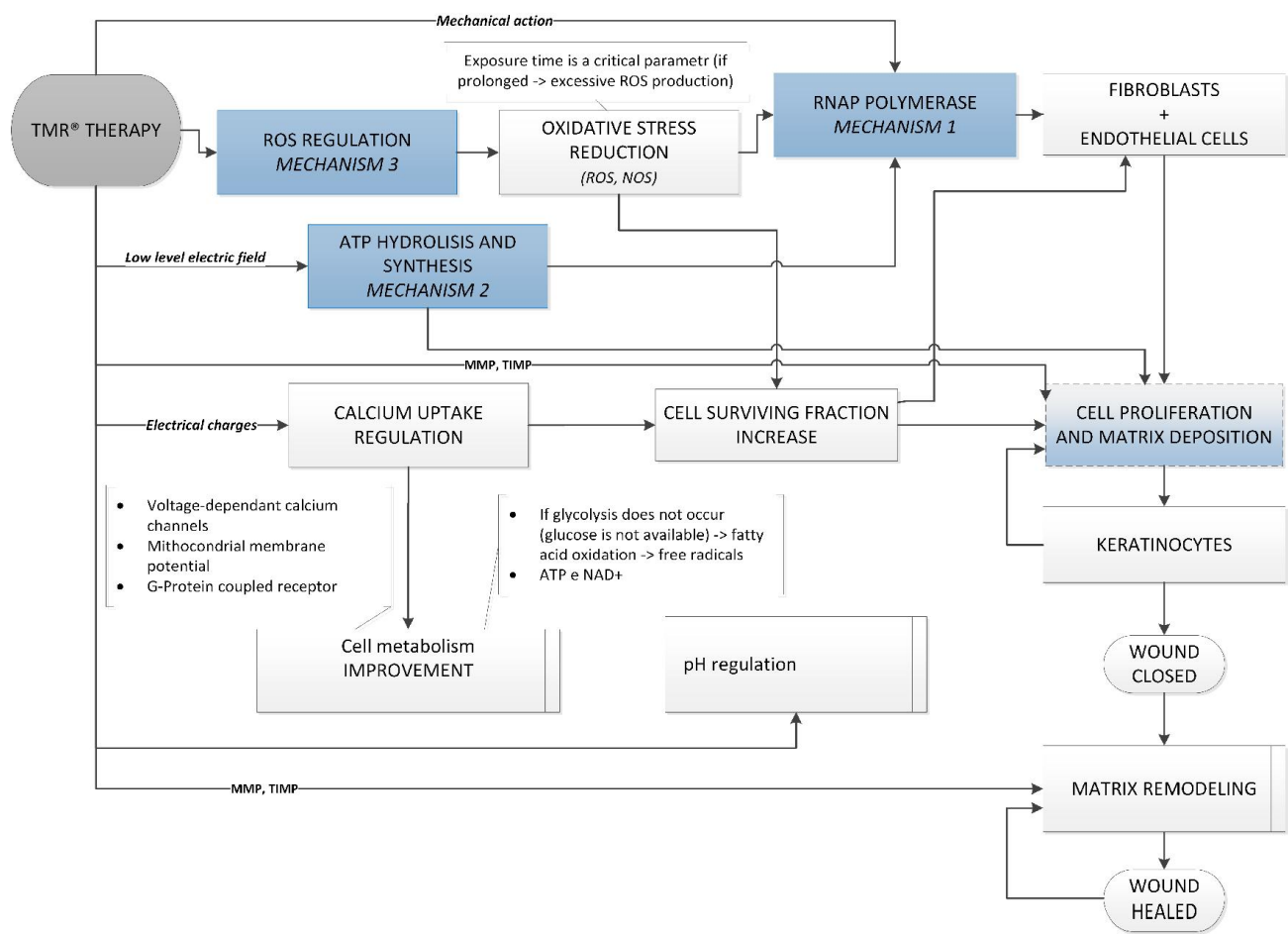

Fig.7. The scheme of the Brizhik-Fermi working principle of the TMR® therapy